\documentclass[iop]{emulateapj}
\usepackage{graphicx,epsfig,fancyhdr,rotating,amsmath,natbib}
\usepackage{txfonts}
\usepackage{epsfig,rotate}
\usepackage{natbib}

\shorttitle{Observation of Recurrent Explosive Events}
\shortauthors{Gupta \& Tripathi}
\begin{document}
\title{IRIS and SDO observation of recurrent explosive events}
\author{G.~R. Gupta \& Durgesh Tripathi}
\affil{Inter-University Centre for Astronomy and Astrophysics, Post Bag-4, Ganeshkhind, Pune 411007, India}
\email{girjesh@iucaa.ernet.in}

\begin{abstract}

Observations of recurrent explosive events (EEs) with time scale of 3-5 minutes are reported. 
These EEs have been observed with the Interface Region Imaging Spectrograph (IRIS) and have a spatial 
dimension of $\sim1.5''$ along the slit. The spectral line profiles of \ion{C}{2}~$1335/1336$ \AA\ and 
\ion{Si}{4}~$1394/1403$ \AA\ become highly broadened both in red as well as
blue wings. Several absorption lines on top of the broadened profiles were 
identified. In addition, emission lines corresponding to neutral lines such as \ion{Cl}{1}~1351.66~{\AA}, 
\ion{C}{1}~1354.29~{\AA}, and \ion{C}{1}~1355.84~{\AA} were identified. 
The \ion{C}{1}~1354.29~{\AA}, and \ion{C}{1}~1355.84 {\AA} lines were found only during 
the EEs whereas \ion{Cl}{1}~1351.66~{\AA} broadens during the EEs.
The estimated lower limit on electron number density obtained using the line ratios of
\ion{Si}{4} and \ion{O}{4} is about $10^{13.5}$ cm$^{-3}$, suggesting that 
the observed events are most likely occurring at heights corresponding to lower chromosphere. To the best 
of our knowledge, for the first time we have detected short-period variability (30 s and 60--90 s) within the 
EE bursts. Observations of photospheric magnetic field underneath EEs indicate 
that negative polarity field emerges in the neighbourhood 
of oppositely directed positive fields which undergo repetitive reconnection (magnetic flux cancellation) events.  
The dynamic changes observed in AIA 1700 \AA, 1600 \AA, \ion{C}{2} 1330 \AA\ and \ion{Si}{4} 1400 \AA\ intensity images corresponded very well  
with the emergence and cancellation of photospheric magnetic field (negative polarity) on the time scale of  
3--5 min. The observations reported here suggests that these EEs are formed due to magnetic reconnection 
and are occurring in the lower chromosphere. 

\end{abstract}


\section{Introduction}\label{intro}

The solar atmosphere is highly dynamic changing on time-scales of minutes to hours. Among many, explosive 
events (EEs) are one of the prominent phenomena observed in the solar transition region. They were discovered 
by \citet{1983ApJ...272..329B} using the observations recorded by the High-Resolution Telescope and Spectrograph 
(HRTS) onboard Black Brant sounding rockets. EEs are characterized with broad line profiles with high velocity
components ($\sim110$ km~s$^{-1}$), which form around 10$^5$ K. They have a spatial scale of $\sim1600$~km 
(2~\arcsec) and life time of $\sim60$~s \citep{1989SoPh..123...41D}.

Using the observations recorded from the Solar Ultraviolet Measurements of Emitted Radiation
\citep[SUMER;][]{1995SoPh..162..189W} spectrograph onboard SOHO, \citet{1997Natur.386..811I} reported 
observations of EEs in chromosphere, which revealed the presence of bi-directional plasma jets as predicted by 
theoretical models of magnetic reconnection. It was also found that these events often occur in bursts lasting 
up to 30~min, whereas individual events may have typical lifetimes of about 1--6 min 
\citep{1997SoPh..175..341I,1998ApJ...497L.109C}. EEs may occur at the same location with period around 
3--5 min and may be triggered by waves found in the solar atmosphere \citep{2004A&A...419.1141N,
2006A&A...446..327D}. It has also been reported that EEs are preferentially located in the regions with weak 
and mixed magnetic polarity \citep{1998ApJ...497L.109C} and are associated with the canceling magnetic flux 
\citep[see e.g.,][]{2008ApJ...687.1398M, 2014ApJ...797...88H}. 

Recently, \citet{2014Sci...346C.315P} studied similar events using the observations recorded by the 
Interface Region Imaging Spectrograph \citep[IRIS; ][]{2014SoPh..289.2733D} and found example of a 
bi-directional jet from the emission profiles of \ion{Si}{4} 1394/1403 \AA\ doublet. They also 
found absorption lines from cooler ions superimposed on these lines suggesting that these hot events are 
occurring in the cooler atmosphere of the Sun. \citet{2014A&A...569L...7S} also found  absorption features 
from a multitude of cool atomic and molecular lines while studying the  broadened profiles of \ion{Si}{4}
transition region lines during the brightening events. \citet{2014ApJ...797...88H} studied single event of EE
along with underlying magnetic field evolution. They found the evidence of magnetic flux cancellation 
and suggested that magnetic reconnection must have taken place during the EE. 

In this paper, we present observations of recurrent EEs using the high-resolution spectroscopic and imaging 
observations from IRIS. We also study the evolution of underlying magnetic field and explore the relationship
between the EEs and the presence of waves in the atmosphere. The paper is organized as follows. We describe the 
observations in section~\ref{obs} and discuss the results in section~\ref{sec:results}. 
We have summarized the results and concluded in section~\ref{conclusion}.

\section{Observations}\label{obs}

\begin{figure*}[htbp]
\centering
\includegraphics[width=14cm]{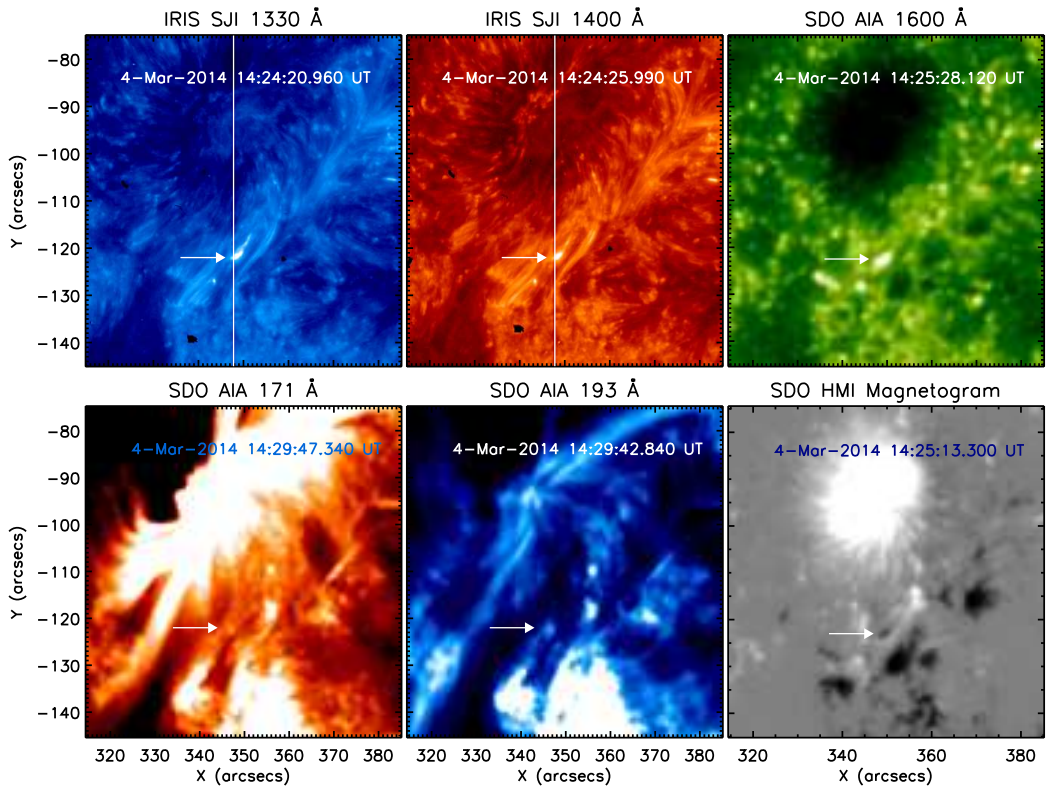}
\caption{The location of EE marked with arrows detected in slit-jaw images of IRIS 1330 {\AA}, and
1400 {\AA}, and AIA 1600 {\AA}, 171 {\AA}, and 193 {\AA} images as labeled. The corresponding HMI magnetic field
map is also shown in bottom right panel. The vertical white continuous line on the top of IRIS slit-jaw
images represents IRIS slit position showing that the slit is passing through the EE studied here.}
\label{fig:context}
\end{figure*}

Data analyzed in this work were obtained by IRIS on 2014 March 4 between 12:39~UT to 14:37~UT using sit-and-stare 
mode. The top left and middle panels of Figure~\ref{fig:context} shows the location of the IRIS slit above the 
slit-jaw images (SJI) obtained in \ion{C}{2}~1330~{\AA} and \ion{Si}{4}~1400 \AA . The top right, bottom left and 
middle panels show corresponding images taken in 1600~{\AA}, 171~{\AA} and 193~{\AA} passbands of the
Atmospheric Imaging Assembly \citep[AIA; ][]{2012SoPh..275...17L} on board the Solar Dynamics Observatory (SDO). 
The bottom right panel displays line-of-sight (LOS) magnetic field measurement obtained from the Helioseismic and 
Magnetic Imager \citep[HMI; ][]{2012SoPh..275..229S}, also on-board SDO. The arrows in all the panels mark the 
location of the recurrent EEs.

IRIS spectra were obtained with an exposure time of 4~s resulting in cadence of approximate 5~s, whereas 
SJI were obtained with an exposure time of 4~s and effective cadence of 15~s. 
In this study, we have used IRIS level-2 data provided by the IRIS team. 
IRIS slit-jaw images from different filters and detectors are already co-aligned for level-2 
data\footnote{\url http://iris.lmsal.com/itn26/calibration.html\#coalignment-between-channels-and-sji-spectra}. 
We also used the fiducial mark along the slit and SJI to correct any offset between them. 
The wobble effect due to thermal flexing between the guide telescope and the main IRIS telescope
is already corrected in regular IRIS operations based on the orbital wobble 
tables \citep[][and private communication, Hui Tian]{2014SoPh..289.2733D}. This correction may still leave an uncertainty of about 1-2 pixels, 
which will not be important for the analysis performed over several spatial pixels.

Data obtained from AIA and HMI have also been utilized in this work. IRIS and AIA observations were co-aligned using IRIS-SJI 
\ion{Si}{4} 1400 \AA\ and AIA 1600 \AA\ images. All the HMI and AIA images obtained in different filters were co-aligned and 
de-rotated with respect to AIA 1600~{\AA} image obtained at 14:09:52~UT using the standard Solar Software 
(SSW) routines. The obtained dataset provide an unique opportunity to study time evolution of recurrent EEs 
using both imaging and spectroscopic observations and to study the evolution of underlying magnetic field.

\section{Results and Discussions}\label{sec:results}

Transition region EEs are identified by very broad emission line profiles showing non-Gaussian enhancements 
in both the wings \citep{1983ApJ...272..329B}. We identified one such small-scale bright structure (with spatial 
width $<1.5~\arcsec$) in the slit-jaw images of \ion{C}{2}~{1330}~{\AA} and \ion{Si}{4}~{1400}{\AA} at the 
location [348.07~\arcsec, -121.44~\arcsec]. The identified EE is marked with arrows in all the panels in 
Figure~\ref{fig:context}. A corresponding animation is provided online, see ee.mp4. An enhancement in intensity 
corresponding to the location and time of EEs identified in \ion{C}{2} 1330 \AA\ and \ion{Si}{4} 1400 \AA\ images is observed in 
AIA 1600~{\AA} image (top right panel). The enhancement can also be identified in images taken in AIA 171~{\AA} and 
AIA 193~{\AA} images, though not as clearly as in AIA 1600~{\AA}. The bottom right panel displays magnetic 
field map. The arrow locates a small-scale ($\approx 2''\times 2''$) negative polarity region (average field 
strength 350 G) surrounded by the positive magnetic polarity regions, which spatially and temporally corresponds 
to the identified EEs. The vertical white line in the IRIS slit-jaw images locates the IRIS slit. Fortuitously, 
the slit was located right at the location where the EE occurred allowing us to perform detailed spectroscopic 
study. Below we discuss the spectroscopy properties of this feature.

\subsection{Evolution of Spectral Line Profiles} \label{sec:spectime}

\begin{figure*}[htbp]
\centering
\includegraphics[width=14cm]{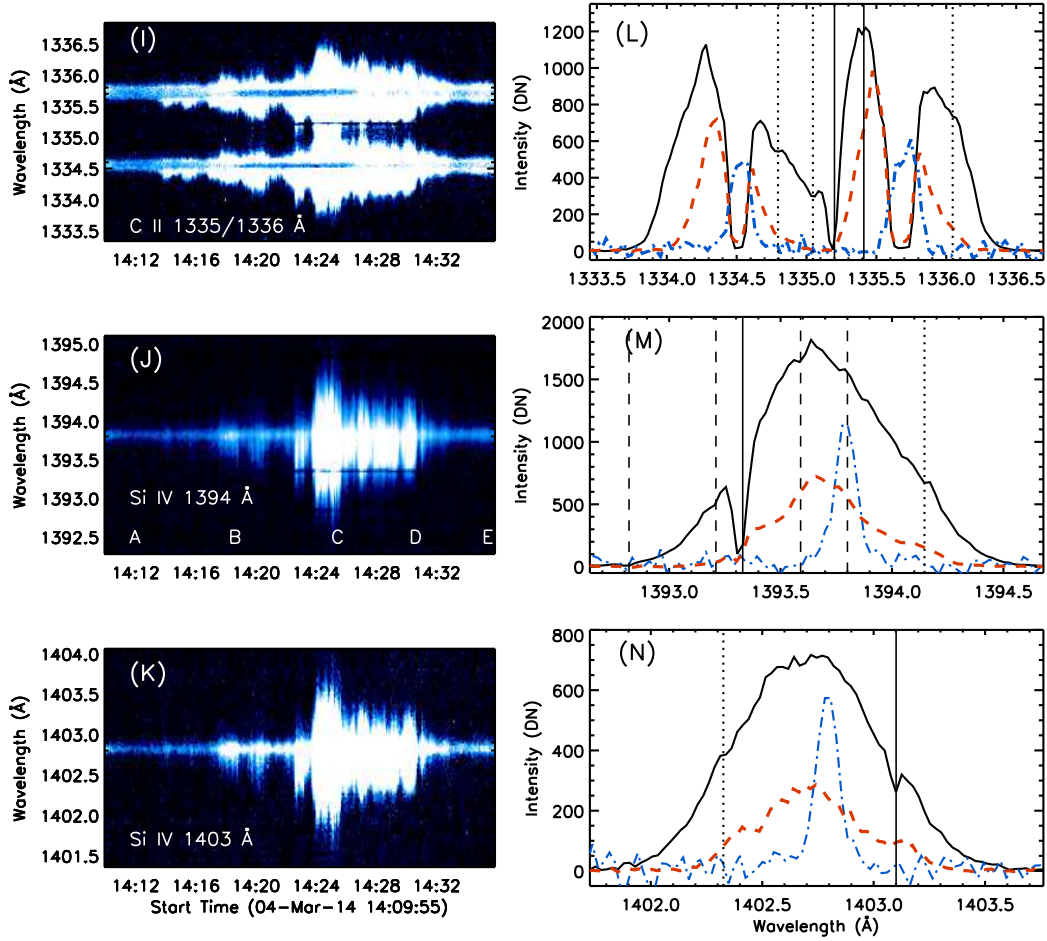}
\caption{Wavelength-time plot of the observed EE in \ion{C}{2}~1335/1336~{\AA} doublet, \ion{Si}{4}~1394~{\AA}, and 
\ion{Si}{4}~1403~{\AA} spectral lines (panels I, J, and K). Typical spectral line profiles at various locations are 
shown in panels L, M, and N, where continuous lines show the line profile at the peak of event (position C in panel J), 
dot-dash line indicates profile during the quiescent time (position A in panel J), and dashed line is for intermediate 
time (position D in panel J). Vertical continuous and dashed lines in panels L, M, and N indicate most prominent and less 
prominent absorption lines from known ions whereas dotted lines indicate absorption features from unknown ions. }
\label{fig:spectra_time}
\end{figure*}

\begin{figure*}[htbp]
\centering
\includegraphics[width=14cm]{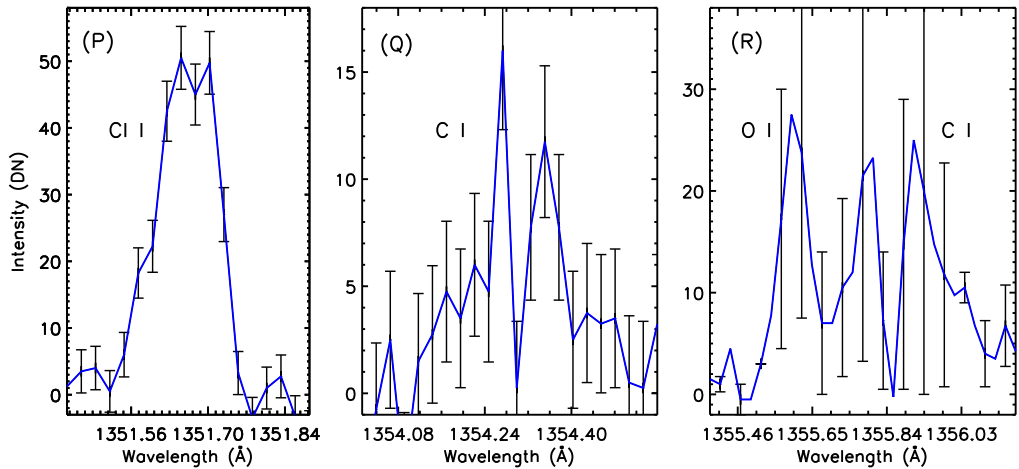}
\caption{Emission lines from neutral atoms such as \ion{Cl}{1}~1351.66~{\AA} (panel P), \ion{C}{1}~1354.29~{\AA} 
(panel Q), and ~1355.84~{\AA} (panel R) observed during EEs corresponding to position C of panel J in 
Figure~\ref{fig:spectra_time}. }
\label{fig:emission}
\end{figure*}

We selected a small portion of IRIS slit data corresponding to the spatial location the EE during the time 
interval of 14:09 UT to 14:36 UT. In Figure~\ref{fig:spectra_time}, we plot time evolution of the spectral line 
profiles of \ion{C}{2}~1335/1336~{\AA} doublet, \ion{Si}{4}~1394~{\AA}, and 1403~{\AA} (see panels I, J and K).
An analysis of the time evolution of the profiles revealed that the EE occurred in multiple bursts with a time 
period of 3--5 min and each burst lasted approximate for 2--3 min. Time evolution of event indicated by the 
enhancement in the intensity and width correspond to each burst (see panels I, J and K).

In the beginning, for the first few bursts, the enhancement in the intensity was by a factor of 4--12 whereas for 
later events the enhancement in intensity was about a factor of 20--40 with respect to the pre-EE phase at the 
same location (see section~\ref{sec:intime}). Since, as can be seen in the animation \textquoteleft ee.mp4\textquoteright ,
the spatial extent and locations of the bursts changed with time, a few bursts in the beginning and at the end could only be 
observed marginally. 
 
Panels L, M and N, respectively, show spectral line profiles of \ion{C}{2} doublet, and two \ion{Si}{4} lines 
(1394~{\AA} and 1403~{\AA}) at different times. The solid lines show the spectra at the peak time of EE 
(labeled as C in panel J) and dashed line show the spectra for the time labeled as D in panel J. For 
comparison, we have also over-plotted the spectra at a quiescent time (labeled as A in panel J) with dashed-
dotted line. As can be inferred from the plots, during quiescent phase \ion{C}{2}~1335/1336~{\AA} 
doublet line show extremely weak self-absorption feature (almost non-existent) in the line center, which could be 
due to large opacity as was pointed out by \citet{2014Sci...346C.315P}. At the times when EEs have started, 
profiles of \ion{C}{2} and \ion{Si}{4} have strongly broadened. This is essentially due to the enhancement in 
both the wings, which were sometimes observed to be asymmetric. During the bursts phase, self-absorption features 
in highly broadened \ion{C}{2} lines increase sharply and bifurcate the spectral line in two (see panel L, solid 
line and dashed line). The \ion{O}{4}~1401~{\AA} spectral line show only about 50\% enhancement in the 
intensity during the peak phase of the EEs, unlike the spectral lines of \ion{C}{2} and \ion{Si}{4}, where the 
enhancement is much higher. 

During the initial times, when EEs were caught marginally by IRIS slit, profiles showed mostly blue-shifted 
emission (or blue wing enhanced). However, during the later time, profiles became very broad with
enhancements in both red and blue wings. We fitted the complete \ion{C}{2} doublet with two positive and two 
negative amplitude Gaussian functions and each \ion{Si}{4} lines with a single Gaussian function.
While fitting, we used error bars, which were provided with IRIS level-2 data on the measured data numbers (DN).
As EEs are believed to have highly non-Gaussian profiles, the reduced $\chi ^2$ and the residuals of the fit were 
generally higher during  the peak activity of the EEs.
We measured Doppler velocity and width (FWHM) of the line with respect to the corresponding average line 
center position obtained at the quiescent phase (at position A labeled in panel J). Doppler width at quiescent 
phase is about 35 km~s$^{-1}$ for both \ion{C}{2} and 25 km~s$^{-1}$ for both \ion{Si}{4} lines.

Both the prominent \ion{Si}{4} lines  show almost identical Doppler velocity and width variation with time.
Doppler velocity and width measured from both the \ion{Si}{4} lines exceeds -50~km~s$^{-1}$ and 
150~km~s$^{-1}$ at the peak of activity around 14:25~UT. However, at the same location and time, \ion{C}{2} 
lines measured Doppler velocity and width exceeding -35~km~s$^{-1}$ and 140~km~s$^{-1}$ at the peak 
activity time, whereas \ion{C}{2} self-absorption lines measured -10~km~s$^{-1}$ and 65~km~s$^{-1}$ 
respectively. Unlike the two \ion{Si}{4} lines, the two \ion{C}{2} emission lines do not show similar Doppler 
velocity and width. \ion{C}{2}~1335.7~{\AA}  emission line consistently shows lower Doppler velocity and width 
as compared to \ion{C}{2}~1334.5~{\AA}. The lower values of Doppler shift and width in 
\ion{C}{2}~1335.7~{\AA} may result due to the presence of strong \ion{Ni}{2}~1335.2~{\AA} absorption line.

From the Figure~\ref{fig:spectra_time}, we identify presence of several absorption features in the emission line 
profiles. The most prominent absorption feature corresponds to \ion{Ni}{2}~1335.20~{\AA} line observed on the 
top of broadened emission line profile of \ion{C}{2}~1335.71~{\AA}. Absorption features of 
\ion{Ni}{2}~1393.33~{\AA}, and \ion{Fe}{2}~1403.10 {\AA} were also observed very prominently on the top of 
profiles of \ion{Si}{4}~1393.76~{\AA}, and 1402.77~{\AA} respectively. These lines are marked with vertical
solid lines. Other less prominent absorption features were also found at 1334.82~{\AA}, 1335.07~{\AA}, 
\ion{Fe}{2}~1335.41~{\AA}, 1336.07~{\AA},  \ion{Fe}{2}~1392.81~{\AA}, \ion{Fe}{2}~1393.21~{\AA}, 
\ion{Fe}{2}~1393.59~{\AA}, \ion{Si}{4}~1393.80~{\AA}, 1394.17~{\AA}, 1402.35~{\AA}, 
\ion{Si}{4}~1402.77~{\AA}. The identified absorption lines are marked with vertical dashed lines whereas 
unidentified lines are marked with vertical dotted lines. Many of these absorption features were reported very 
recently by \citet{2014A&A...569L...7S,2014Sci...346C.315P} and \citet{2015arXiv150105706Y}. Some of them
have also reported the presence of self-absorption lines on the top of broadened profiles of \ion{C}{2}~1334.54~{\AA} 
and 1335.71~{\AA}, \ion{Si}{4}~1393.76~{\AA} and 1402.77~{\AA} during the explosive events.

We have also identified several emission lines coming from neutral atoms such as 
\ion{Cl}{1}~1351.66~{\AA}, \ion{C}{1}~1354.29~{\AA}, \ion{C}{1}~1355.84~{\AA} and \ion{O}{1}~1355.60 {\AA}. 
All these lines except the \ion{O}{1}~1355.60 {\AA} show presence of either self-absorption or absorption features 
(see Figure~\ref{fig:emission}) during the peak activity time of explosive events. To the best of our knowledge, this is 
first report of appearance of self-absorption features in the emission lines of neutral atoms. However, we note that the 
associated error bars on the data points are relatively large. Therefore, a more detailed investigation is required to 
confirm that these are indeed self absorption features. The \ion{C}{1}~1354.29~{\AA}, and 1355.84~{\AA} lines were 
not present during the quiescent phase and were detected only during the EE, whereas 
\ion{Cl}{1}~1351.66~{\AA} line was present during the quiescent phase and broadened during the EEs. The 
\ion{O}{1}~1355.6 {\AA} was present both during the quiescent phase as well as during EEs. However, 
surprisingly, we didn't find any absorption feature and noticeable line broadening of \ion{O}{1}~1355.6 {\AA} line 
during any time. such as \ion{Cl}{1}~1351.66~{\AA}, \ion{C}{1}~1354.29~{\AA}, and
\ion{C}{1}~1355.84~{\AA} during the EEs  in the IRIS spectra.

\subsection{Intensity Evolution of Explosive Events} \label{sec:intime}

\begin{figure*}[htbp]
\centering
\includegraphics[width=12cm]{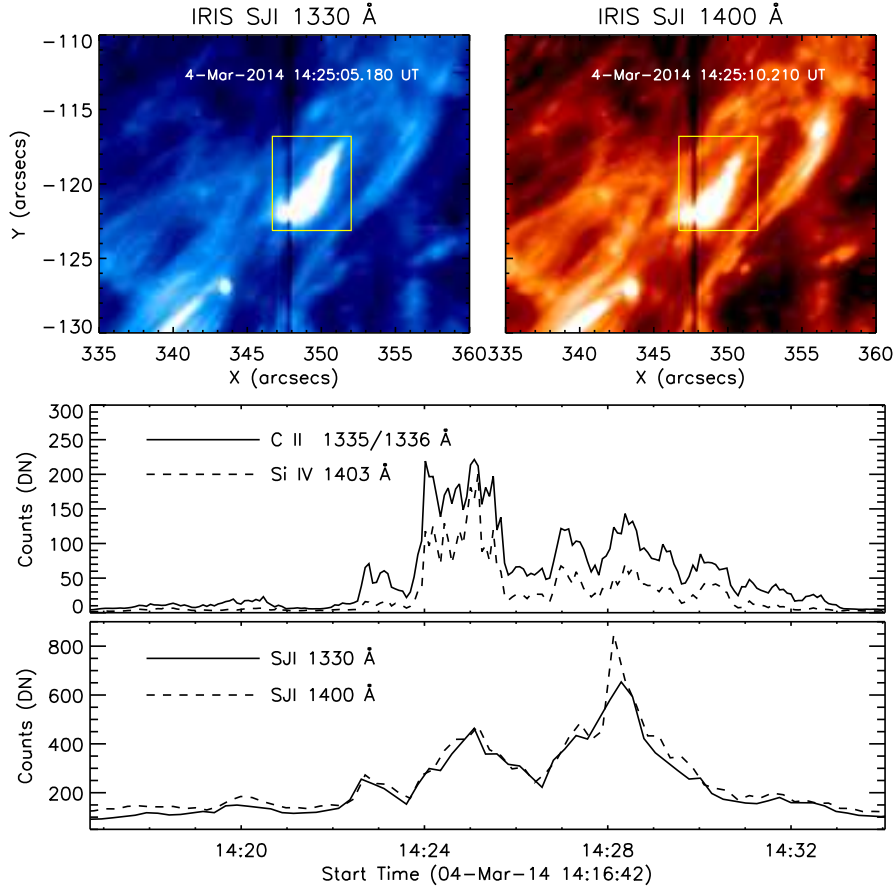}
\caption{Top panels: Area chosen to obtain intensity evolution in IRIS 1330~{\AA} and 1400~{\AA} passbands. Middle 
panel: Intensity variation with time as recorded by IRIS spectrometer in \ion{C}{2}~1335/1336~{\AA} doublet,
and \ion{Si}{4}~1403~{\AA} spectral line. Bottom panel: Intensity
variation with time as obtained from IRIS 1330~{\AA} and 1400~{\AA} images in the chosen area.}
\label{fig:iris_lc}
\end{figure*}

IRIS records both spectroscopic and imaging data of the solar atmosphere simultaneously. While spectra recorded with 
the slit has very limited field of view, IRIS slit-jaw imager (SJI) can record data in the bigger field of view. As 
mentioned in the Section~\ref{sec:spectime}, IRIS slit caught few bursts marginally. Thus we also use SJI 
images obtained in 1330~{\AA} and 1400~{\AA} passbands of IRIS to study the full time evolution of the whole 
EEs sequence. We also looked into the different AIA passbands to find any signatures of these events in the 
upper layers of the atmosphere. 

In top panels of Figure~\ref{fig:iris_lc} (see movie $ee\_iris.mp4$), we show an area chosen to study the intensity 
evolution in IRIS 1330~{\AA} and 1400~{\AA} passbands. Middle panel of Figure~\ref{fig:iris_lc} displays the change 
in the intensity with time as recorded by IRIS spectrometer. The light curves clearly reveal that the first and last
few bursts were weak. The strongest burst were seen starting at around 14:24~UT. In the bottom panel of 
Figure~\ref{fig:iris_lc}, we plot intensity obtained in the chosen area with time in different passbands of IRIS.
Recently, \citet{2015ApJ...803...44M} found that IRIS passbands have significant contributions from  the continuum.
However, continuum effect will be significant only when signals in the \ion{C}{2} and \ion{Si}{4} lines are too weak to explain 
the presence of observed features in the 1300 and 1400~{\AA} SJIs. In the case of EEs, signals in \ion{C}{2} and \ion{Si}{4}
profiles are very strong (stronger by a factor of 4-40), which are seen in the SJIs, thus, the contribution from the continuum 
may be ignored. The light curve for two IRIS passbands show very similar evolutionary characteristics. The light curve obtained 
from IRIS passbands indicate that total intensity of EEs was increasing with time in the beginning and started 
decreasing only after 14:28~UT. The strongest burst in SJI images were seen at around 14:28~UT, which is 
partially being captured by the IRIS slit. It should also be noted that since the IRIS spectra were recorded at higher cadence 
as compared with the SJI images, we are able to see even shorter bursts within each EE recorded by SJI.

\begin{figure}[htbp]
\centering
\includegraphics[width=7cm,angle=90]{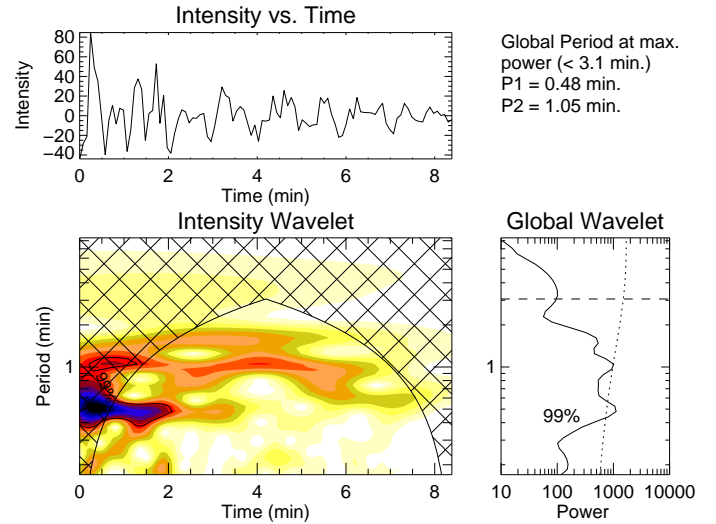}
\caption{Wavelet analysis result for the intensity variation with time (starting from 14:21:28 UT) as recorded by 
IRIS \ion{C}{2}~1335/1336~{\AA} spectral lines. The top panel show the  intensity variation with time. The 
bottom-left panel show the color-inverted wavelet power spectrum with 99\% confidence-level contours, while the 
bottom-right panel show the global wavelet power spectrum with 99\% global confidence level drawn. The periods P1 
and P2 at the locations of the first two maxima in the global wavelet spectrum are shown above the global wavelet 
spectrum.}
\label{fig:wavelet1}
\end{figure}

\begin{figure}[htbp]
\centering
\includegraphics[width=7cm,angle=90]{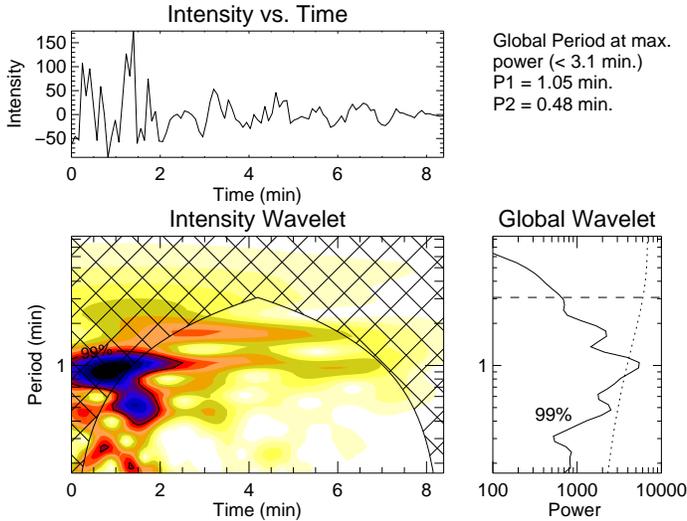} 
\caption{Same as Figure~\ref{fig:wavelet1} but for \ion{Si}{4}~1394~{\AA} spectral line.}
\label{fig:wavelet2}
\end{figure}

To find the period of intensity variability within the EE, we performed wavelet analysis \citep{1998BAMS...79...61T} 
on the intensity variation with time as recorded by \ion{C}{2} 1335/1336~{\AA} (see Figure~\ref{fig:wavelet1}) and
\ion{Si}{4}~1394~{\AA} and 1403~{\AA} (see Figure~\ref{fig:wavelet2}) spectral lines. The top panels in both the figures show the 
variation of intensity with time starting at 14:21:28~UT. The bottom left panels are wavelet power 
spectrum (color inverted) with 99 \% confidence levels and bottom right panels are global wavelet power spectrum 
(wavelet power spectrum summed over time) with 99\% global confidence. We used background (trend) subtracted intensity to 
obtain the wavelet power spectrum. Background (trend) was obtained by taking the 10-point running average of the 
intensity variation. The power spectra obtained for both the spectral lines reveal the presence of short-period intensity variability in 
the EE light curves with two distinct periods of around 30~s and 60--90~s in addition to 3--5~min variability. 
We also find that even without background (trend) subtraction, the obtained power peaks at the same periods in the wavelet spectra, 
though with lower confidence level. We note here that plots shown here is created for the intensities which were obtained 
after taking summation over the spectral profile range of the lines. However, we have found that the variability obtained after 
performing wavelet analysis on Gaussian amplitudes, and product of Gaussian width and amplitude are essentially the same. 
Further, we rule out the effect of wobble, if any, on the obtained short term variabilities. It has been found that during a course of half orbit 
($\approx48$ min), IRIS pointing moves by about 17 {--}18 pixels ($\approx 3\arcsec$,  in course of full orbit, pointing may come back to its original position)
on the Sun \citep{2014SoPh..289.2733D}. This means that slit location will change by about 1~pixel in 2.8 min, 
whereas, in this analysis, we found intensity variabilities  of the order of 1 min and less within the individual EEs extended over several 
spatial pixels. Therefore,  instrumental wobble do not have affected our analysis and henceforth the obtained short-period intensity variabilities.

Repetitive nature of explosive events and jets with periodicities of 3--5~min have been reported earlier 
\citep[see e.g,][]{2004A&A...419.1141N,2006A&A...446..327D} and have been attributed to presence 
of MHD waves in the atmosphere with similar periods. However, to the best of our knowledge, this is the first time a 
short-period variability  with periods of 30~s and 60--90~s has been detected in EEs. Recently, using Hi-C data 
\citet{2015arXiv150106507P} found evidence of short-period (30~s and 53--73~s) oscillations in braided magnetic 
region, providing evidence in favor of connection between short period waves and bursts. However, at this point 
we can not conclude whether oscillations are driving burst events or vice-versa.

\subsection{Electron Density in Explosive Events} \label{sec:density}

\begin{figure*}[htbp]
\centering
\includegraphics[width=12cm]{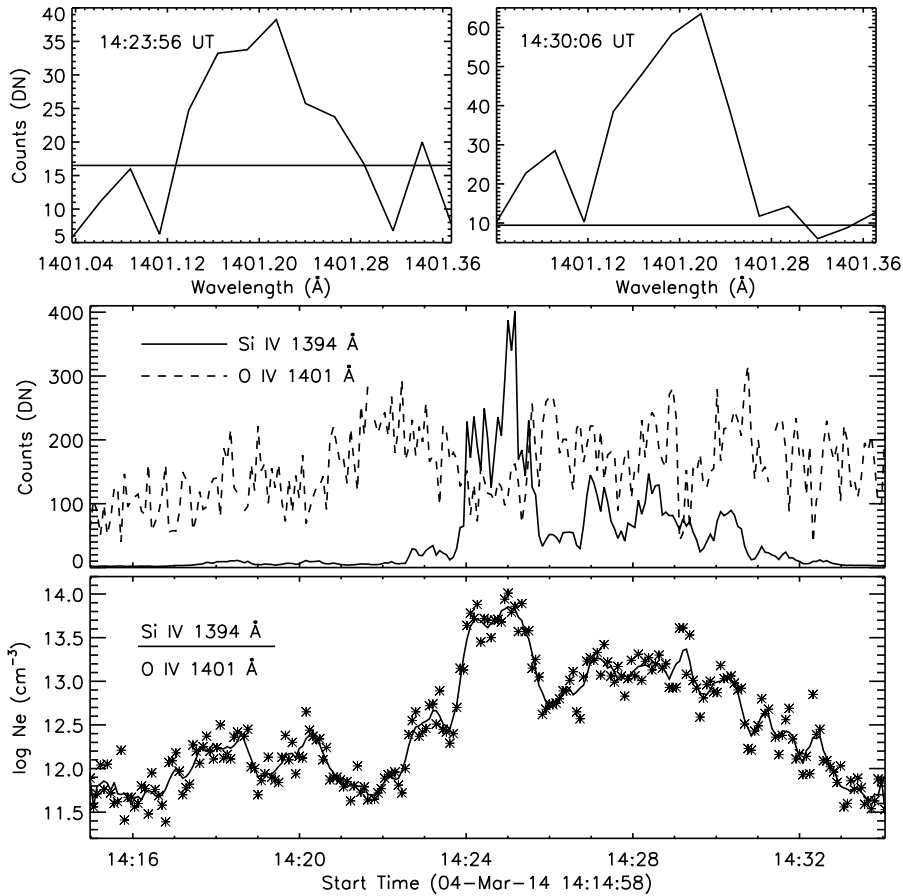}
\caption{Top panel: \ion{O}{4}~1401~{\AA} profiles obtained at two different instants of time after binning over 
7~pixels along the slit at the EE location. Middle panel: Intensity variation of \ion{O}{4}~1401~{\AA} and 
\ion{Si}{4}~1393.76~{\AA} lines with time. Bottom panel: Temporal evolution of estimated electron number density 
(lower limit) obtained from the ratios of \ion{Si}{4} to \ion{O}{4} lines.}
\label{fig:oiv_dens}
\end{figure*}

To estimate electron density during the EEs, density sensitive \ion{O}{4}~1401~{\AA} and 1399~{\AA} line pair
may be used, although see \citep{2014ApJ...780L..12D}. Both these lines are very weak in the IRIS  spectra. By 
binning over a few pixels the signal to noise ratio (SNR) for \ion{O}{4}~1401~{\AA} can be improved. However, 
this does not improve the SNR for \ion{O}{4}~1399~{\AA}. 

Based on the suggestions provided by \citet{2014Sci...346C.315P}, we estimated electron density using 
\ion{O}{4}~1401.16~{\AA} and \ion{Si}{4}~1393.76~{\AA} \citep[see however ][]{2013A&A...557L...9D}, which 
only provides a lower limit of the electron density. Since our goal is to get an order-of-magnitude estimate
for the electron density, method suggested by \citet{2014Sci...346C.315P}
will serve the purpose. For this calculation, we have used the photospheric abundances 
of \citet{1998SSRv...85..161G} and CHIANTI ionization equilibrium \citep{2012ApJ...744...99L}. 
 
In order to improve the signal to noise ratio, we binned spectra over 7 pixel along the direction of the slit. 
The top two panels in Figure~\ref{fig:oiv_dens} show two examples of \ion{O}{4}~1401~{\AA} line profiles at two 
different times. As can be seen from the line profiles there are still not enough counts to allow to fit a Gaussian, 
we summed over the profile (wavelength range from 1400.94~{\AA} to 1401.50~{\AA}) and subtracted the contribution
of continuum to estimate the \ion{O}{4}~1401~{\AA} contribution. Although the \ion{Si}{4}~1393.76~{\AA} line is strong 
enough to fit a Gaussian and obtain the Gaussian integrated intensity, however, for consistency we have used the same method to 
obtain the intensity as for \ion{O}{4}~1401~{\AA}.

The resultant intensity variation of both the lines are plotted in middle panel of Figure~\ref{fig:oiv_dens}.
Electron number densities estimate during the quiescent time is around $\sim10^{11.7}$ cm$^{-3}$ which increases
to $\sim10^{13.5}$ cm$^{-3}$ during the peak of the activity (see bottom panel of Figure~\ref{fig:oiv_dens}).
We note that since the intensities derived for \ion{O}{4}~1401~{\AA} is the upper limit, the densities obtained
here are essentially the lower limit. This estimated lower limit on the electron number density together with 
appearance of neutral lines (\ion{C}{1}~1354.29~{\AA}, and \ion{C}{1}~1355.84~{\AA}) during the EE strongly suggests 
that event probably occurred somewhere in the lower chromosphere, as was also pointed out by 
\citet{2014Sci...346C.315P}

\subsection{Comparison with AIA Observations} \label{sec:aia}

\begin{figure*}[htbp]
\centering
\includegraphics[width=14cm]{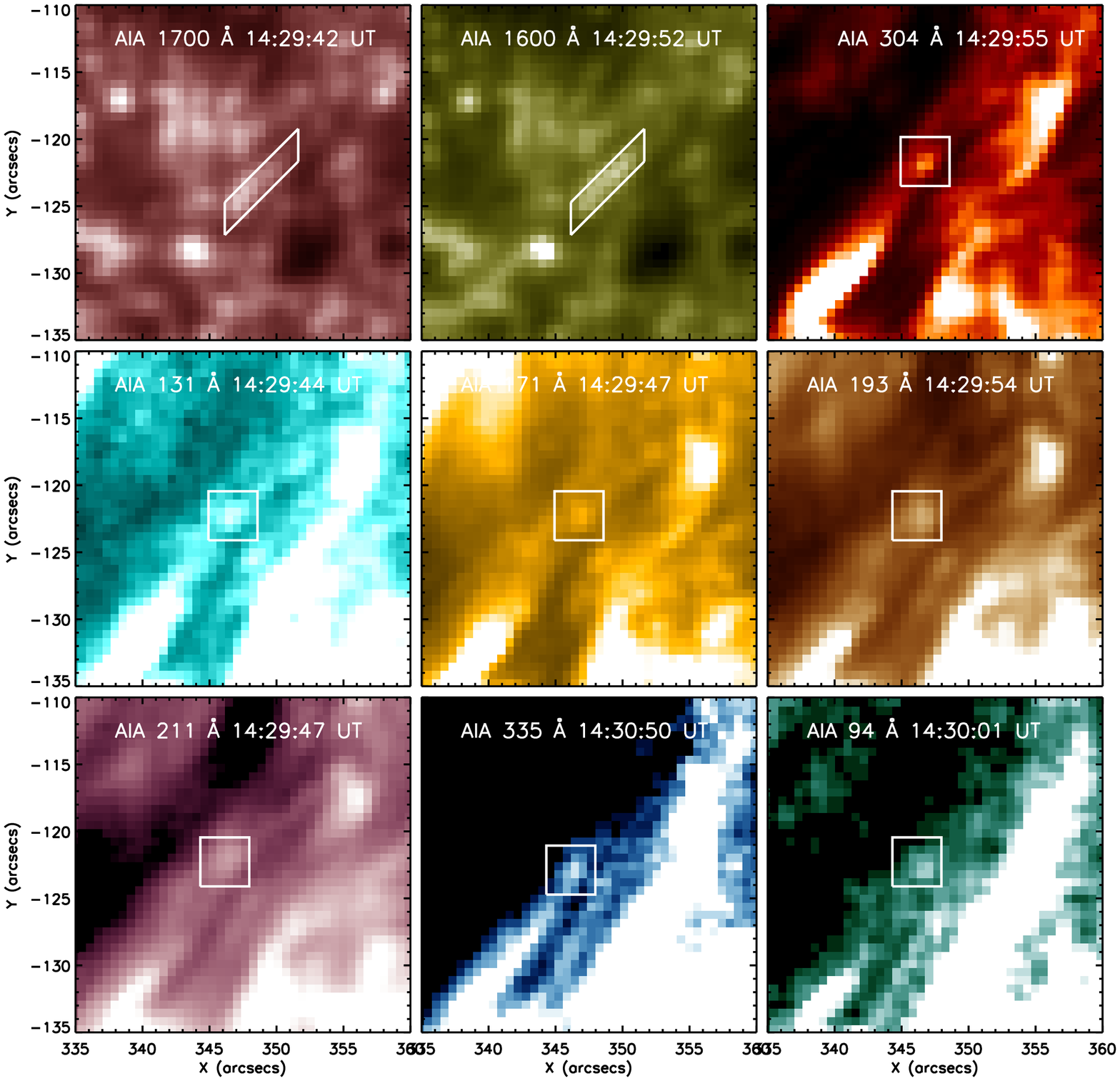}
\caption{Images obtained from AIA 1700~{\AA}, 1600~{\AA}, 304~{\AA}, 131~{\AA}, 171~{\AA}, 193~{\AA}, 211~{\AA}, 335~{\AA}, and 94~{\AA}
passbands as labeled during the EE.}
\label{fig:aia_img}
\end{figure*}

\begin{figure}[htbp]
\centering
\includegraphics[width=9cm]{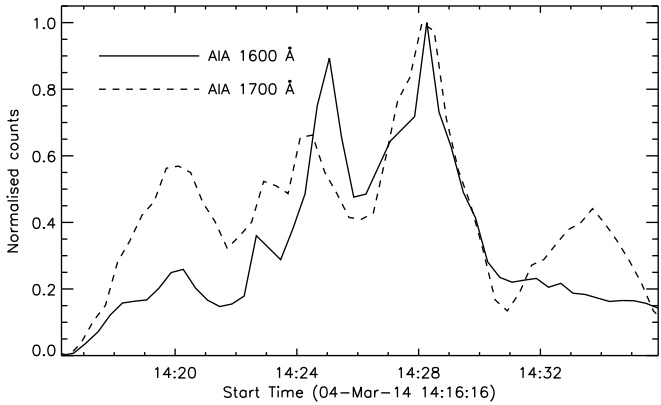}
\caption{Intensity light curves obtained from AIA 1700~{\AA}, and 1600~{\AA}   passbands as labeled at the location of EEs.}
\label{fig:aia_lc}
\end{figure}

To study if there were any hotter counterparts of the repetitive EEs, we examined the different AIA passbands. 
Figure~\ref{fig:aia_img} displays AIA images taken just after the  peak at 14:28 UT as seen in the bottom panel 
of Figure~\ref{fig:iris_lc} in all its passbands. For complete time evolution, see movie $ee\_aia.mp4$. The observations
recorded using 1700~{\AA} and 1600~{\AA} passband show appearance and movement of EEs in full extent -- 
marked by a rectangular box -- similar to that seen in IRIS-SJI images, see the light curves plotted in Figure~\ref{fig:aia_lc}. 
The light curves are plotted using normalized intensities, which were obtained as $[(I(t)-min (I(t))=max(I(t)-min(I(t))]$.
However, the hotter channels of AIA show only a small but distinguishable brightening that is spatially correlated to that 
of the EEs. These brightening are located in the images using a square. The full extent of these brightening is not clear 
in these hotter channels. Also, their is a time lag of about 5 minutes in the appearance of these brightening in hotter 
channels. This makes us wonder if these brightening are similar to those observed using SJI as well as AIA 1700~{\AA} 
and 1600~{\AA} images. 
We would also like to point out that coronal loops that connect to more southerly negative flux from the sunspot is more likely crossing the field of view.
As pointed out by \citet{2014Sci...346C.315P}, EUV radiation is strongly attenuated by overlying  hydrogen so it would not be surprising to see no signature
of the events in AIA passbands at the same time. However, as time progress more energy is released, resulting in ionization of hydrogen with time, we see
response in AIA passbands for later events. 

\begin{figure}[htbp]
\centering
\includegraphics[width=9cm]{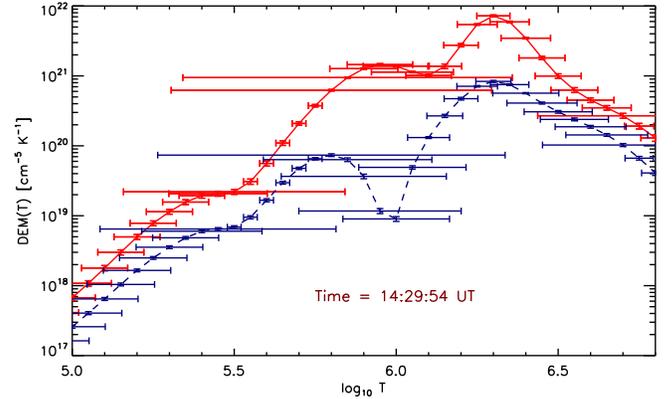}
\caption{DEM curve obtained using coronal emission data from AIA 131~{\AA}, 171~{\AA}, 193~{\AA}, 211~{\AA}, 
335~{\AA}, and 94~{\AA} passbands at the spatial location of EE and at nearby time of 14:29:54 UT. Continuous 
line red curve is obtained using the original intensity whereas dashed line blue curve is obtained after subtracting 
background.}
\label{fig:aia_dem}
\end{figure}

In order to understand the nature of the small brightening seen in the AIA hotter channels, we performed differential 
emission measure (DEM) analysis to obtain the differential emission measure distribution using the regularized inversion 
method of \citet{2012A&A...539A.146H}. The DEM was obtained using  AIA  131~{\AA}, 171~{\AA}, 193~{\AA}, 211~{\AA}, 
335~{\AA}, and 94~{\AA} intensities recorded at nearby time of 14:29:54~UT when emission in AIA 193~{\AA} channel peaks.
The obtained DEM curve is plotted in Figure~\ref{fig:aia_dem}. The red curve is DEM before background subtraction and 
the blue curve is after the background subtraction.
Background intensities are  intensities obtained at the EE location just before the brightening start appearing in AIA images.
 As expected, the overall DEM values decreases after the background subtraction. The DEM curve
show rather a strong dip at $\sim$~1 MK after the background subtraction. However, the peak emission is still coming from
a temperature of $\log\,T=6.3$. This suggests that the brightening seen in the AIA channels is at coronal temperature. 
There is also another peak at  $\log\,T=5.8$. However, the error bars are much bigger at this location in the plot. 

If the brightening seen in hotter AIA channels are related to those seen in IRIS-SJI as well as  AIA 1600~{\AA} 
and 1700~{\AA} images, then the plasma must be heated to about 2 MK.
As mentioned earlier, the time analysis shows that there is a lag about 5 minutes in the appearance of these brightening in 
hotter channels. Also we note that by the time these brightening appear in hotter channels, they disappear from the SJI 
images. However, at this point, we are not able to conclude if these brightening are exactly related to each other.

\subsection{Magnetic Field Evolution of Explosive Event}
\label{sec:magtime}

\begin{figure*}[htbp]
\centering
\includegraphics[width=12cm]{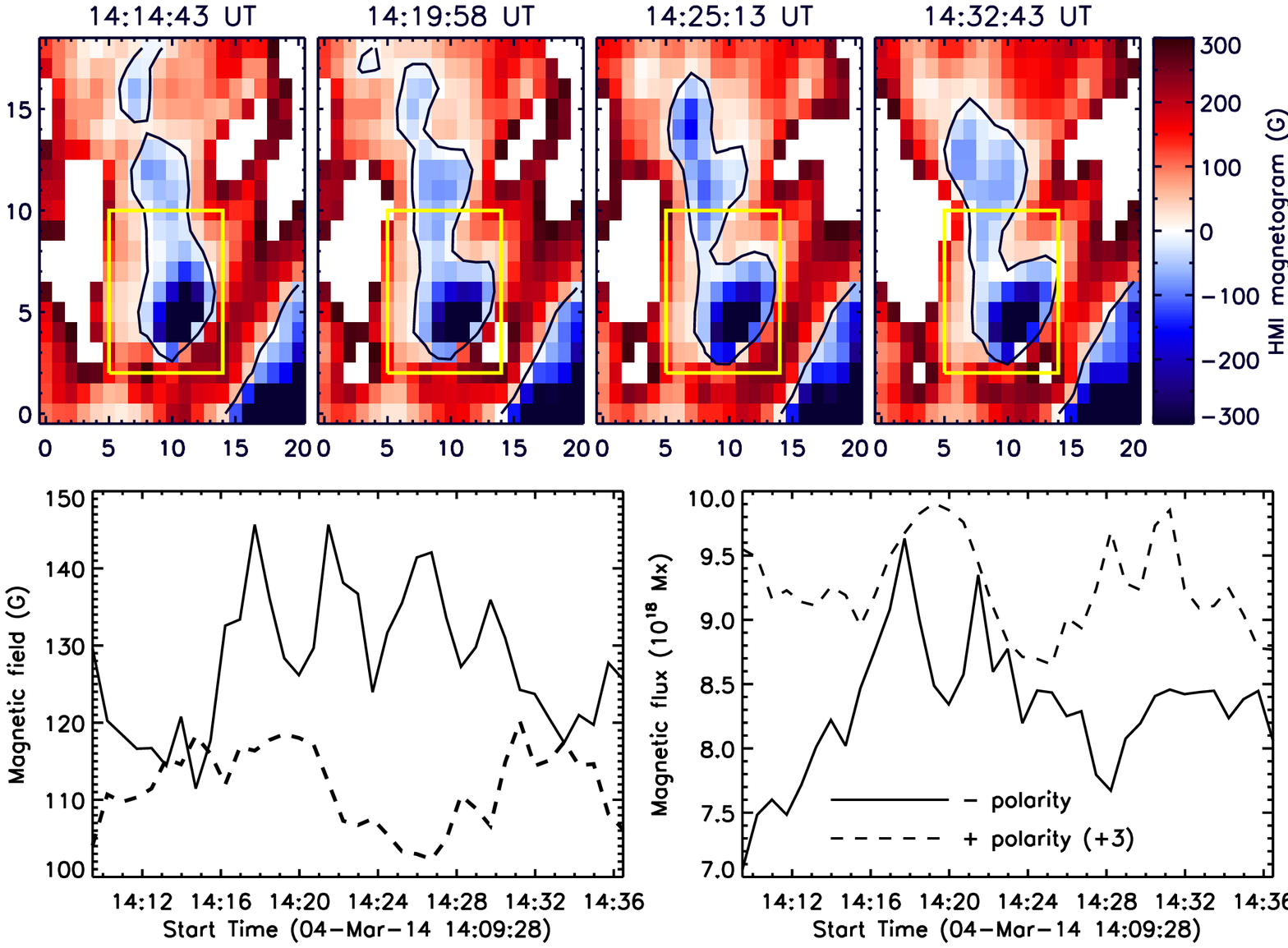}
\caption{Time evolution of underlying photospheric magnetic field of EE as recorded by HMI magnetogram.
Top panels show the small region where EEs were observed. Contour levels are over-plotted at the field
strength of 0 G. Over-plotted box indicate area chosen to estimate average field strength and flux at the EE source region.
Bottom panels provide the time evolution of positive (dashed line) and negative (continuous line) 
field strength (left), and flux (right).}
\label{fig:ee_mag}
\end{figure*}

\begin{figure*}[htbp]
\centering
\includegraphics[width=12cm]{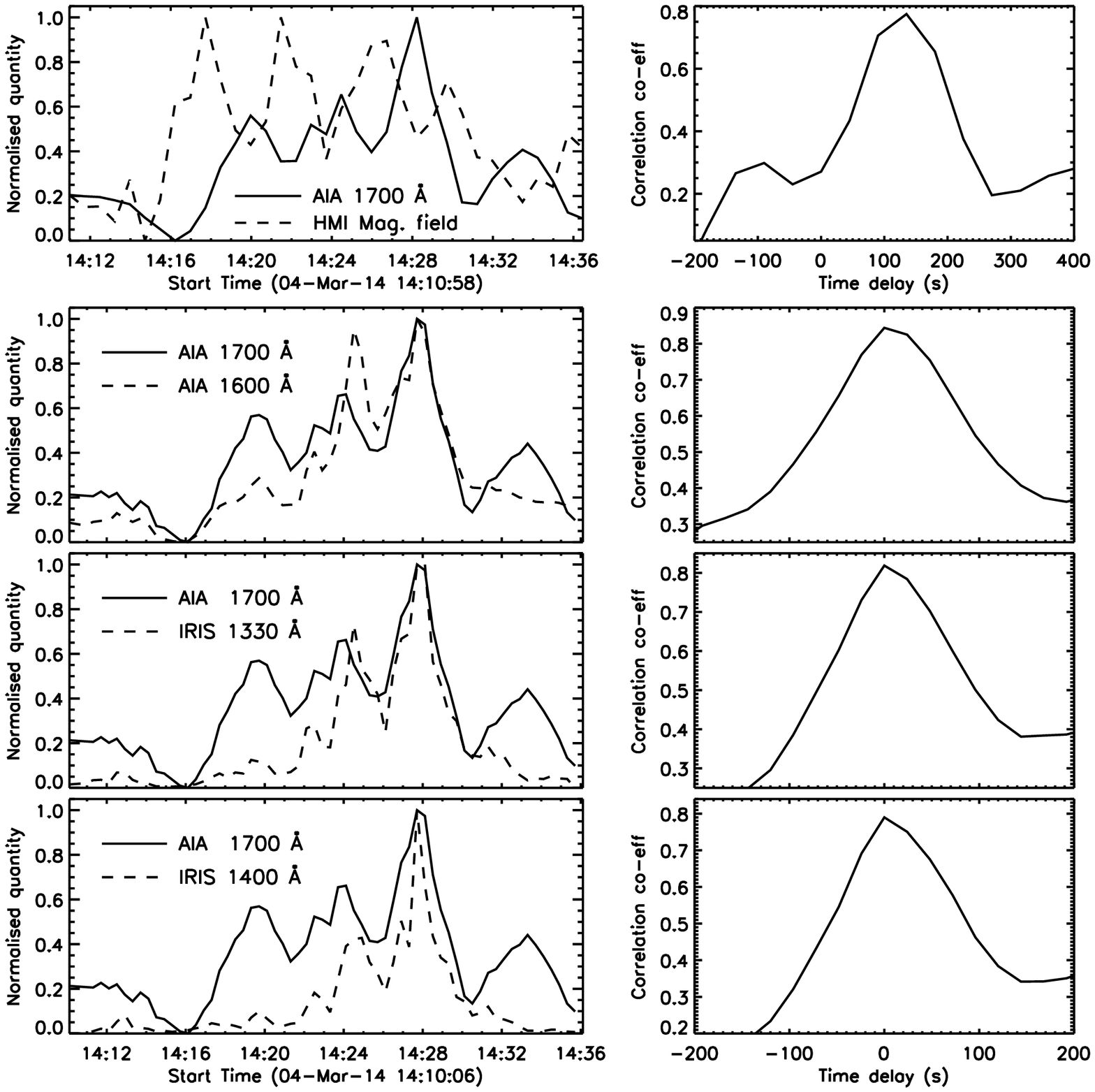}
\caption{Cross-correlation analysis between intensity evolution of EEs obtained from AIA 1700 \AA\
(continuous line, all panels) with HMI magnetogram (negative polarity, top panel, dashed line), AIA 1600 \AA\
(second panel, dashed line), IRIS 1330 \AA\ (third panel, dashed line), and IRIS 1400 \AA\
(bottom panel, dashed line). Cross-correlation co-efficient values obtained with original signal at
different time delays are plotted in the right panels for the respective light curve pairs.}
\label{fig:tdelay}
\end{figure*}

We analyzed HMI/SDO magnetogram data to study evolution of photospheric magnetic field during the EE. 
Figure~\ref{fig:ee_mag} (see movie $ee\_hmi.mp4$) shows presence of small-scale negative field (enclosed with a
rectangular box) surrounded by the positive field at the location of EE. We have studied the temporal evolution of 
positive and negative magnetic field as well as flux in the the boxed region and plotted in Figure~\ref{fig:ee_mag}. 
As can be seen from the plots, negative polarity field and flux (solid lines in bottom panels) 
increases and decreases during the EEs on the time scale of 3--5 min. Whereas positive polarity field 
and flux (dashed lines in bottom panels) show somewhat decay during that time. 

To find any correlation between fluctuations in magnetic field measured from HMI and in intensity during EE measured from 
AIA and IRIS, we perform correlation analysis at different time delays. HMI magnetogram is obtained at photospheric height 
whereas AIA 1700~{\AA} is formed at photosphere and temperature minimum height, thus, are more suitable for direct 
correlation study. Correlation of intensity fluctuations obtained from AIA 1600~{\AA}, IRIS~1330~{\AA}, and 1400~{\AA}  are
performed with respect to that obtained from AIA~1700~{\AA}. As cadence of HMI is 45~s whereas that of AIA~1600~{\AA}, 
1700~{\AA} and IRIS~1330~{\AA}, 1400~{\AA} are 24~s and 15~s respectively. We interpolated AIA~1700~{\AA} with respect 
to HMI time sequence using spline routines. We also interpolated AIA 1600~{\AA}, and IRIS 1330~{\AA}, 1400~{\AA} light 
curves with respect to time sequence of original AIA 1700~{\AA}. Thus, we obtained cross-correlation coefficients at different 
time delays for the pairs of HMI magnetogram (negative polarity) and AIA 1700~{\AA} light curve, AIA 1700~{\AA} and 
1600~{\AA}, AIA~1700~{\AA} and  IRIS~1330~{\AA}, and AIA~1700~{\AA} and IRIS~1400~{\AA} light curves. We plot 
respective pairs of light curves and correlation coefficients at different time delays in Figure~\ref{fig:tdelay}. Maximum 
correlation value between HMI magnetogram (negative polarity) and AIA 1700~{\AA} light curve is about 0.77 at the time 
delay of 135~s, which corresponds to 3 HMI time frames. However, that between AIA 1700--1600~{\AA}, AIA~1700--IRIS~1330~{\AA},
AIA~1700--IRIS~1400~{\AA} pairs are about 0.84, 0.82, 0.79 respectively at the time delay of 0 s. Maximum 
cross-correlation co-efficient between AIA 1600~{\AA} and IRIS 1330~{\AA} and 1400~{\AA}  are above 0.90 at time delay 
of 0 s (not shown here). Thus, the obtained results suggest that intensity fluctuations as recorded from AIA 1600~{\AA} and 
1700~{\AA} and IRIS passbands are connected to fluctuations in photospheric magnetic fields as recorded from HMI with the 
time delay of about 135 s. Increase in flux and field could be related to flux emergence, whereas decrease in magnetic field 
and flux could be associated with magnetic flux cancellation resulting due to magnetic reconnection. 
Magnetic flux cancellation events underneath EEs had been previously reported by \citet{1998ApJ...497L.109C} 
and recently by \citet{2014ApJ...797...88H}. However, to the best of our knowledge, this is first 
time we report such a correlated change in AIA 1700 \AA, 1600 \AA, \ion{C}{2} 1330 \AA , and \ion{Si}{4} 1400 \AA\ 
intensity with photospheric magnetic field (in this case negative polarity) on the time scale of 3{--}5 min during the recurrent bursts 
of EEs.

In order to get further insight of the magnetic field evolution, we performed a long term study of the sunspot 
region i.e. from the time when it first emerged on the east limb on Feb 25, 2014 \textsl{AR 11990} till the time of the analyzed 
EEs using HMI observations. The region appeared as a simple sunspot (with negative polarity) within which an emergence of a positive 
field region was detected. This positive field region evolved with time and developed as a complete sunspot with positive polarity, 
suggesting a highly complex field evolution and formation of a delta-sunspot (Gupta et al. in preparation). With the evolution of the 
sunspot, various moving magnetic features (MMFs) were observed. Therefore, in the context, it is plausible to conclude that the active 
region is relatively a young with several small-scale MMFs around the sunspot. The EEs studied in current paper was related to one 
of the MMFs.

\section{Summary and Conclusions} \label{conclusion}

In this paper we studied observations of recurrent EEs using simultaneous spectroscopic and imaging observations 
recorded by IRIS, AIA images and HMI magnetogram. To the best of our knowledge, this is first report of such 
recurrent EEs using IRIS data. The recurring time scale of these EEs were about 3-5 min. During the event, line 
profiles of \ion{C}{2} and \ion{Si}{4} showed enhanced broadening with Doppler velocity and width exceeding more 
than -50 km~s$^{-1}$ and 150 km~s$^{-1}$ respectively. In addition, we identified several absorption lines on the 
top of broadened emission lines of \ion{C}{2} and \ion{Si}{4}. Moreover, we also found a few neutral atom lines such 
as \ion{Cl}{1}~1351.66~{\AA}, \ion{C}{1}~1354.29~{\AA}, and \ion{C}{1}~1355.84~{\AA} with possible self-
absorption features. While \ion{C}{1}~1354.29~{\AA},1355.84~{\AA} lines appeared only during the EEs, the 
\ion{Cl}{1}~1351.66~{\AA}, which was present before the EEs, showed broadening during the EEs. The lower limit on 
electron densities obtained using the method proposed by \citet{2014Sci...346C.315P} was about $10^{13.5}$ cm$^{-3}$. 
Using the high cadence spectroscopic observations we also discovered short period variability ($\sim$ 30 s and 60--90 s) 
within the EE bursts. 

The analysis the LOS of photospheric magnetic field measured by HMI underneath explosive events indicated
emergence as well as cancellation of magnetic flux. The negative polarity magnetic flux showed continuous increase 
with a periodic fluctuations, suggesting localized cancellations (see bottom panels of Figure~\ref{fig:ee_mag}). The 
changes in AIA 1700 \AA, 1600 \AA, \ion{C}{2} 1330 \AA , and \ion{Si}{4} 1400 \AA\  intensities correlated extremely well with the changes in the negative polarity 
magnetic flux on the time scale of  3--5 min during the recurrent bursts of explosive events, which has not been reported 
earlier, to the best of our knowledge.

The observations of strong broadening in \ion{C}{2} and \ion{Si}{4} spectral lines along with self-absorption and 
estimated high electron density suggests that these are lower chromospheric features. The correlation with canceling 
magnetic features supports the idea of formation of these features due to magnetic reconnection. We believe that
these features can be explained by resistive emergence of magnetic flux underneath the photosphere and 
expanding in the upper layers \citep[see e.g.,][]{2004ApJ...614.1099P, 2007ApJ...657L..53I}. This concept 
has successfully explained the formation of Ellerman Bombs, which were originally detected in H$\alpha$ 
observations \citep[][]{1917ApJ....46..298E}.  In this scenario,  a number of $\Omega$ loops rise due to
well known Parker's instability from underneath the photosphere and expand. While emergence and expansion, 
they interact with each other and the process of magnetic reconnection occurs. The process of reconnection may 
produce the EEs and heat the plasma locally. This process may result in a cool material being stacked upon locally 
heated material, thus resulting in a cool absorption lines superimposed on the hot emission lines as proposed 
by  \citet{2014Sci...346C.315P}. Therefore, the observed EEs are most likely the upper atmospheric 
counterpart of Ellerman Bombs. This concept, however, needs to be throughly verified using 
observations recorded in H$\alpha$ line core and wings along with IRIS observations and forward modeling of the
spectral lines using MHD simulations of realistic solar atmosphere.

\acknowledgments
We thank the referee for the insightful comments, which improved the quality of the manuscript.
GRG is supported through the INSPIRE Faculty Award of the Department of Science and Technology (DST), India. 
DT acknowledges the support from the Max-Planck Partner Group on Coupling and Dynamics of the Solar Atmosphere
at IUCAA. IRIS is a NASA small explorer mission developed and operated by LMSAL with mission operations executed at
NASA Ames Research center and major contributions to downlink communications funded by the Norwegian 
Space Center (NSC, Norway) through an ESA PRODEX contract. AIA and HMI data are courtesy of SDO (NASA).
Facilities: SDO (AIA, HMI). CHIANTI is a collaborative project involving George Mason University, the University of Michigan (USA) and the University of Cambridge (UK).

\end{document}